\documentclass[aps,prb,reprint,superscriptaddress,amssymb,showpacs,showkeys]{revtex4-1}

\usepackage{graphicx,xspace,xcolor}

\newcommand{\Bk}{B$_{\mathrm{k}}$\xspace}
\newcommand{\uu}[1]{\ensuremath\,\mathrm{#1}}

\begin{document}

\date{\today}
\title{Towards predictive modelling of near-edge structures in electron energy loss spectra of AlN based ternary alloys}

\author{D.~Holec}
\email{david.holec@unileoben.ac.at}
\affiliation{Department of Physical Metallurgy and Materials Testing, Montanuniversit\"at Leoben, Franz-Josef-Strasse 18, A-8700 Leoben, Austria}
\affiliation{Department of Materials Science and Metallurgy, University of Cambridge, Pembroke Street,Cambridge CB2 3QZ, United Kingdom}
\author{R.~Rachbauer}
\affiliation{Department of Physical Metallurgy and Materials Testing, Montanuniversit\"at Leoben, Franz-Josef-Strasse 18, A-8700 Leoben, Austria}
\author{D.~Kiener}
\affiliation{Erich Schmid Institute of Materials Science, Austrian Academy of Sciences, Jahnstrasse 12, A-8700 Leoben, Austria}
\affiliation{Department of Materials Physics, Montanuniversit\"at Leoben, Franz-Josef-Strasse 18, A-8700 Leoben, Austria}
\author{P.D.~Cherns}
\affiliation{Department of Materials Science and Metallurgy, University of Cambridge, Pembroke Street,Cambridge CB2 3QZ, United Kingdom}
\author{P.M.F.J.~Costa}
\affiliation{Department of Materials Science and Metallurgy, University of Cambridge, Pembroke Street,Cambridge CB2 3QZ, United Kingdom}
\affiliation{CICECO, Department of Ceramics and Glass Engineering, University of Aveiro, 3810-193 Aveiro, Portugal}
\author{C.~McAleese}
\affiliation{Department of Materials Science and Metallurgy, University of Cambridge, Pembroke Street,Cambridge CB2 3QZ, United Kingdom}
\author{P.H.~Mayrhofer}
\affiliation{Department of Physical Metallurgy and Materials Testing, Montanuniversit\"at Leoben, Franz-Josef-Strasse 18, A-8700 Leoben, Austria}
\author{C.J.~Humphreys}
\affiliation{Department of Materials Science and Metallurgy, University of Cambridge, Pembroke Street,Cambridge CB2 3QZ, United Kingdom}

\begin{abstract}
Although electron energy loss near edge structure analysis provides a tool for experimentally probing unoccupied density of states, a detailed comparison with simulations is necessary in order to understand the origin of individual peaks. This paper presents a {\color{purple}density functional theory based} technique for predicting the N K-edge for ternary (quasi-binary) nitrogen alloys by adopting a core hole approach, a methodology that has been successful for binary nitride compounds. It is demonstrated that using the spectra of binary compounds for optimising the core hole charge ($0.35\uu{e}$ for cubic Ti$_{1-x}$Al$_x$N and $0.45\uu{e}$ for wurtzite Al$_x$Ga$_{1-x}$N), the predicted spectra evolutions of the ternary alloys agree well with the experiments. The spectral features are subsequently discussed in terms of the electronic structure and bonding of the alloys.
\end{abstract}



\maketitle

\section{Introduction}

Wurtzite aluminium nitride (w-AlN), gallium nitride (w-GaN) and their ternary alloy w-Al$_x$Ga$_{1-x}$N are important materials for devices such as light emitting diodes (LEDs) and laser diodes (LDs) \citep{jain00}. Cubic titanium nitride (c-TiN) and in particular its alloy with AlN, cubic Ti$_{1-x}$Al$_x$N, are widely used hard coatings due to their high hardness, corrosion and oxidation resistance \citep{mayrhofer06b}. In both ternary alloys, a crucial requirement for getting the optimal application-tailored properties is an accurate control of both the composition \emph{and} structure of the alloy. 

% Besides X-ray diffraction (XRD) and transmission electron microscopy (TEM) techniques such as energy dispersive X-ray spectroscopy (EDX), $Z$-contrast scanning TEM or geometrical phase analysis, one may also use 
Electron energy loss spectroscopy (EELS) is a powerful technique to microanalyse compositional features \citep{egerton96}. The spectra can be recorded with very high spatial resolution, thus taking advantage of the high sensitivity of EELS to local changes in the electronic structure of materials \citep{botton96, rez08}. Its subset, electron energy loss near edge structure (ELNES) reflects the density of unoccupied states and provides thus an experimental probe for this part of the electronic structure of materials. However, the interaction of high energy electrons with lattice atoms does not always have a straightforward interpretation. In order to understand the experimental data, measured EELS spectra need to be compared in detail with their calculated counterparts.

{\color{purple}X-ray absorbtion near-edge structure (XANES) is a closely related technique to ELNES. The edge shapes are very similar \citep{keast01, rez08}, it also allows for studying polarisation effects \citep{katsikini98}, and it has been successfully applied to diluted alloys \citep{ciatto05}. Therefore, the here proposed methodology is highly relevant also for XANES.}

Recent developments of theoretical methods for solid state physics have provided the EELS community with increasingly reliable and comprehensive tools to simulate ELNES. A particularly important example is the Telnes program which is distributed as a part of Wien2k \citep{wien2k}, an all-electron full-potential linearised augmented plane-wave (FP-LAPW) code. It has been suggested in the literature \citep[see, e.g.][]{hebert07} that a calculation including a core hole provides a better description of the excitation process by means of the standard (ground state) density functional theory (DFT). To create a core hole, one takes an electron (or a fraction of it) from its ground state position (N $1s$-state in the case of N K-edge) and puts it in the lowest unoccupied state above the Fermi level (N $2p$) or adds it as a background charge to keep the total charge of the cell neutral \citep{hebert07}. This can be easily realised in Wien2k where all electrons are accessible and explicitly treated. Recently, the core hole calculations have become routinely available also for pseudopotential codes \citep{seabourne09, mizoguchi09}.

A considerable effort has been spent on studying ELNES of binary nitrides with respect to their crystal structures (wurtzite, rock salt, zinc blende) \citep{lawniczak-jablonska97,lawniczak-jablonska00,mizoguchi03, mizoguchi04, sennour03, gao04}, polarisation effects \citep{lawniczak-jablonska97, lawniczak-jablonska00, keast02, radtke03, gao04}, doping \citep{serin98}, and stoichiometry \citep{mirguet06, lazar08, kwak10}. The literature is vast and provides a good background for understanding the origin of peaks in ELNES in terms of bonding, and thus establishes a solid basis for a finger-print identification of materials and their properties.

Nonetheless, there are only a few reports on the compositional dependence of ELNES of ternary alloys. \citet{keast03} measured N K-edge ELNES of In$_x$Ga$_{1-x}$N alloys and correlated their findings with DFT calculations using, however, extremely small supercells. \citet{mackenzie05} used ordered structures and averaging of the boundary binaries spectra to get a guess on the shape of Ti$_{1-x}$Al$_x$N N K-edge. \citet{holec08} showed that using small ordered cells failed to reproduce the experimental spectrum of an Al$_x$Ga$_{1-x}$N alloy. To date, a systematic study showing the effect of alloying on the ELNES shape (``evolution'' of the edge), as well as discussing the computational methodology in a recipe-like form, is missing. The present paper is aiming to fill this gap by facilitating a {\em semi-empirical approach} {\color{purple}in which} the calculation parameters (e.g. the core hole charge) are first {\em adjusted} to reproduce the spectra of the boundary binary systems, and subsequently used to {\em predict} the N K-edge evolution of ternary alloys.

\section{Methodology}

\subsection{Calculation details}

\begin{figure}
  \centering
  \includegraphics[width=8cm]{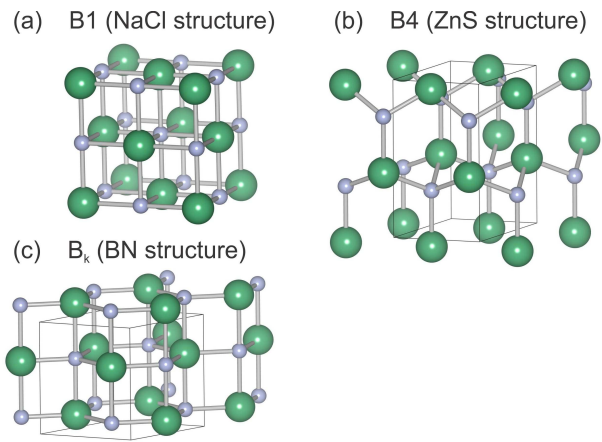}
  \caption{Crystal structures investigated in this work: (a) cubic B1 (NaCl prototype), (b) wurtzite B4 (ZnS prototype), and (c) five-coordinated hexagonal \Bk structure (BN prototype).}
  \label{fig:structures}
\end{figure}

The individual structures are modelled with supercells constructed using a special quasi-random structure (SQS) approach \citep{wei90}. All alloys considered in this paper are quasi-binary which means that mixing of elements (either Ti and Al or Al and Ga) takes place only on one sublattice (bigger atoms in Fig.~\ref{fig:structures}); the other sublattice is fully occupied with N atoms. $3\times3\times2$ (36 atoms) and $2\times2\times2$ (32 atoms) supercells were used for the cubic B1 and wurtzite B4 modifications, respectively. The short range order parameters were optimised for pairs up to the fourth order, triplets up to the third order and quadruplets up to the second order \citep{wei90, holec10-diss}. {\color{purple}More details about the cells and the process of their generation can be found in Ref.~\onlinecite{holec10-diss}.}

The cubic and wurtzite structures of the Ti$_{1-x}$Al$_x$N alloy were optimised with respect to volume (lattice constants) as well as to internal (local) relaxations, for which we used the  VASP code \citep{vasp2, vasp1} together with the projector augmented wave pseudopotentials \citep{vasp-pot} employing the generalised gradient approximation (GGA) as parametrised by \cite{wang91}. We used $500\uu{eV}$ for the plane-wave cut-off energy and a minimum of $\approx600\,k$-points$\cdot$atom (usually more). Such parameters guarantee the calculation accuracy in the order of meV/atom. The obtained equilibrium lattice parameters (see Tab.~\ref{tab:lattice}) and energies are in good agreement with those previously published \citep{mayrhofer06a, alling07, alling08}.

\begin{table}
  \centering
  \begin{ruledtabular}
  \begin{tabular}{l|c|cc|c}
     & structure & $a\ [\mbox{\AA}]$ & $c\ [\mbox{\AA}]$ & $B_0\ [\mbox{GPa}]$ \\ \hline%\hline
    TiN & B1 & 4.256 &       & 292 \\
        & \Bk & 3.524 & 4.227 & 238 \\% \hline
    AlN & B1 & 4.070 &       & 253 \\ 
        & B4 & 3.129 & 5.016 & 197 \\% \hline
    GaN & B4 & 3.216 & 5.238 & 169
  \end{tabular}
  \end{ruledtabular}
  \caption{Optimised lattice parameters and bulk moduli of the binary compounds used in this work.}
  \label{tab:lattice}
\end{table}

TiN is mechanically unstable in the four co-ordinated wurtzite (B4) structure (Fig.~\ref{fig:structures}b) and relaxes into a five-coordinated \Bk structure (Fig.~\ref{fig:structures}c). The reason for this is that the presence of $d$ electrons favours a different hybridisation scheme ($sp^3d$) than the tetrahedral $sp^3$ (see Refs. \onlinecite{atkins98, holec08}). Consequently, the wurtzite variant of the Ti$_{1-x}$Al$_x$N alloy becomes unstable around $x\approx0.6$ whereas for $x\lessapprox0.6$ the \Bk structure is obtained \citep{holec08, holec_inprep}. This is, however, the composition where also the phase transition to a lower energy, experimentally observed cubic variant happens \citep{mayrhofer06a}. Therefore, in the following we present only results for the high Al containing w-Ti$_{1-x}$Al$_x$N.

As for the Al$_x$Ga$_{1-x}$N alloy, we optimised the crystal lattices only for the boundary binary compounds (Tab.~\ref{tab:lattice}), AlN and GaN, and then used Vegard's rule to obtain the lattice constants for the intermediate composition. This is justified by the work of \citet{Dridi03} who showed for this alloy that the lattice parameters (unlike the band gap) exhibit a linear dependence on the composition.

Electronic properties and the ELNES spectra were calculated using the Wien2k code \citep{wien2k} employing the GGA--PBE parametrisation \citep{perdew96} of the exchange-correlation potential. An equivalent of $\approx900$ $k$-points within the whole first Brillouin zone of the unit cell, the expansion of the spherical harmonics up to $l=10$ inside the non-overlapping muffin tin (MT) spheres, and $R_{\mathrm{MT}}k_{\max}=7$  were used\footnote{Those are the standard parameters. However, we checked their values for convergence.}. The  MT radii were automatically set by structGen (a part of the Wien2k package) to values $\approx1.70$--$1.80$, $\approx1.95$--$2.00$, $\approx1.85$--$1.95$ and $\approx1.90$--$1.95\uu{a.u.}$ for N, Ti, Al, and Ga atom, respectively. The spin polarisation effects were not taken into account. The core holes were implemented by reducing the N $1s$ core level occupation on a specific site and putting the corresponding charge in the background in order to keep the cell neutral \citep{hebert07}. ELNES was calculated using the Telnes program, a part of Wien2k. {\color{purple}This was repeated for all N sites in the supercell. The spectrum representing the particular alloy composition was obtained by averaging this set of N K-edges.}

\subsection{Experiment}

In order to experimentally confirm the \textit{ab initio} predicted ELNES spectra, two material systems, Ti$_{1-x}$Al$_x$N and Al$_x$Ga$_{1-x}$N were investigated. 

The Ti$_{1-x}$Al$_x$N samples were deposited in Leoben using the plasma-assisted unbalanced magnetron sputtering technique \citep[see][]{rachbauer10}. The variation of the Al mole fraction $x$ in Ti$_{1-x}$Al$_x$N was achieved by using powder metallurgically produced targets (PLANSEE AG, 99\% purity), with Ti/Al ratios of 1, 0.5 and 0.33, and manual placing of additional Ti or Al platelets ($\varnothing 5\times3\uu{mm}$) in the racetrack of the targets, respectively. TEM sample preparation was performed by Ar-ion thinning in a Gatan precision ion polishing system (PIPS) at $4$ and $2.2\uu{keV}$ in plan view. The EELS measurements were carried out on a Cs corrected Jeol 2100F operated at $200\uu{kV}$ and equipped with a Gatan Tridiem GIF camera using nanobeam diffraction mode. This ensures high signal to noise ratios and makes it possible to acquire information from individual grains, necessary for the investigation of the polycrystalline Ti$_{1-x}$Al$_x$N films. Thus, several different grains were measured for each alloy composition to rule out possible orientation effects. The spectra were recorded with a dispersion of $0.3\uu{eV/channel}$ and the energy resolution, measured by the full-width at half-maximum of the zero-loss peak, was $1.5$--$1.8\uu{eV}$. The convergence and collection semi-angles during analysis were $5\uu{mrad}$ and $>8\uu{mrad}$, respectively. 

The Al$_x$Ga$_{1-x}$N films were epitaxially grown at Cambridge by $6\times2$-inch Thomas Swan Close-Coupled Showerhead metalorganic vapour-phase epitaxy (MOVPE) system \citep{mcaleese04}. A standard two-step growth method was used to deposit a $5\uu{\mu m}$ thick GaN pseudo-substrate on $(0001)$ sapphire \citep{datta04} on which the Al$_x$Ga$_{1-x}$N layers were grown at $1020\uu{^\circ C}$. The different compositions were obtained by varying the flow rate for Al and Ga precursors. The EELS spectra were obtained on a FEI Tecnai F20 microscope equipped with a Schottky FEG source, Gatan Imaging Filter and operated at $200\uu{kV}$. {\color{purple}The electron beam was parallel to the $\langle0001\rangle$ direction.}

Prior to the comparison with the \textit{ab initio} calculations, all measured spectra were corrected for the dark current and the channel-to-channel gain variation. The pre-edge background was extrapolated using a power-law function and subtracted from the original data \citep{egerton96}.

\section{Results}

\subsection{Binary compounds}\label{binaries}

\begin{figure*}
  \centering
  \includegraphics{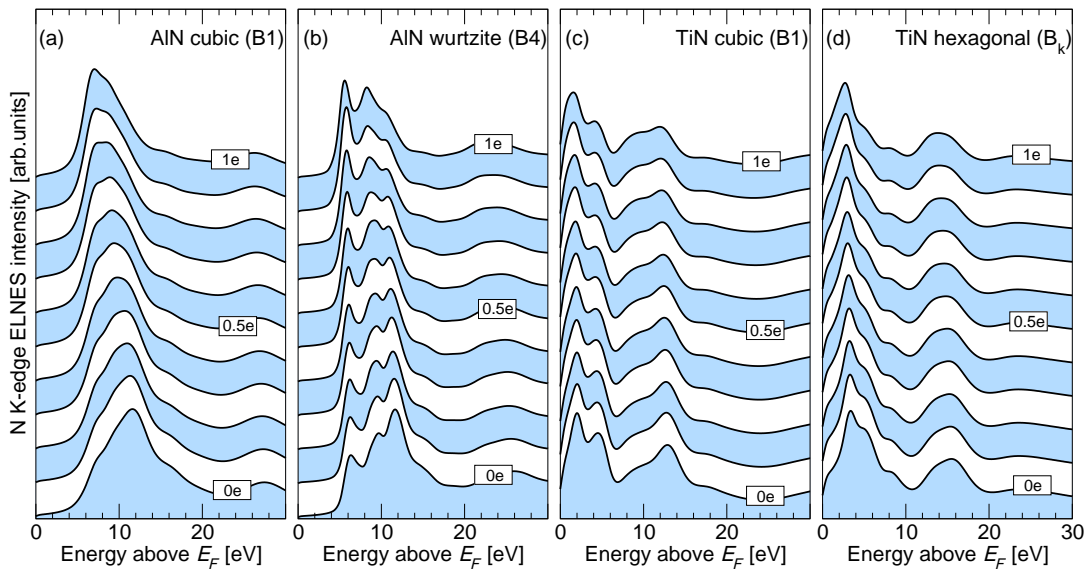}
  \caption{Calculated N K-edge ELNES as a function of the core hole charge changing from a ground state calculation ($0\uu{e}$) to a full core hole ($1\uu{e}$): (a) cubic B1 AlN, (b) wurtzite B4 AlN, (c) cubic B1 TiN, and (d) hexagonal \Bk TiN.}
  \label{fig:binaries}
\end{figure*}

The strategy adopted in this paper is to find calculation parameters that reproduce the N K-edge ELNES for the binary compounds as closely as possible, and subsequently use these settings for {\em predicting} the ELNES evolution of the alloys. Figure~\ref{fig:binaries} shows how the edge shape of cubic and wurtzite/hexagonal AlN and TiN changes with increasing core hole charge from $0\uu{e}$ (ground state) to $1\uu{e}$ (final excited state -- full core hole). The effect is stronger for AlN than for TiN. This is due to fast core hole screening in TiN originating from its metallic character \citep{rez08}. Nevertheless, some small changes can still be observed, e.g. the peak broadening for the cubic modification or the disappearing high-energy shoulder of the main peak for the hexagonal TiN with increasing core hole charge. \citet{lazar08} arrived at the same conclusion for c-TiN based on the comparison of their calculations with experimental measurements.

\begin{figure}
  \centering
  \includegraphics{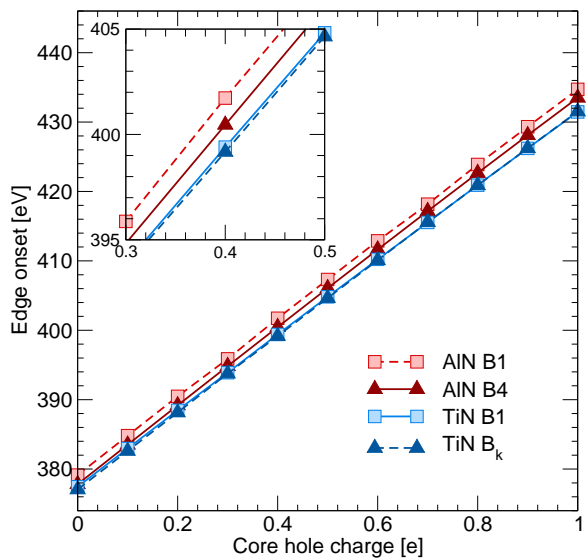}
  \caption{N K-edge onset (energy difference between the initial core state and lowest unoccupied state) as a function of the core hole charge for the two allotropes of each AlN and TiN. The inset shows zoomed-in region around the experimentally observed N K-edge onset energy.}
  \label{fig:edge-onset}
\end{figure}

The strong effect of the core hole charge on the AlN N K-edge shape has been discussed in the literature \citep{mizoguchi09, holec08}. A detailed analysis of the relative peak positions and intensities for the wurtzite AlN revealed that a core hole charge $\approx0.5$--$0.6\uu{e}$ reproduces the experimental ELNES the best \citep{holec10-diss}. Such a comparison could still be misleading as the experimental results also depend strongly on the acquisition conditions \citep{radtke03}. Since the spectra around $0.5\uu{e}$  (Slater's transition state) {\color{purple}for each allotrope look akin} and are almost equally resembling the experiment, another criterion was adopted here. The edge onset, measured as the energy between the initial core state and the lowest unoccupied state \cite{rez08}, is plotted as a function of the core hole charge in Fig.~\ref{fig:edge-onset}. This dependence is rather strong, and using the experimental value for the edge onset allows an optimal value of the core hole charge to be estimated. Taking $402\uu{eV}$ for w-AlN \citep[or this work]{mizoguchi03, mizoguchi09a} and $397\uu{eV}$ for c-TiN \citep[or this work]{rashkova07, craven95} yields $0.45\uu{e}$ and $0.35\uu{e}$, respectively, which are the values used in this work.

\subsection{Ti$_{1-x}$Al$_x$N alloy}

\begin{figure*}
  \centering
  \includegraphics{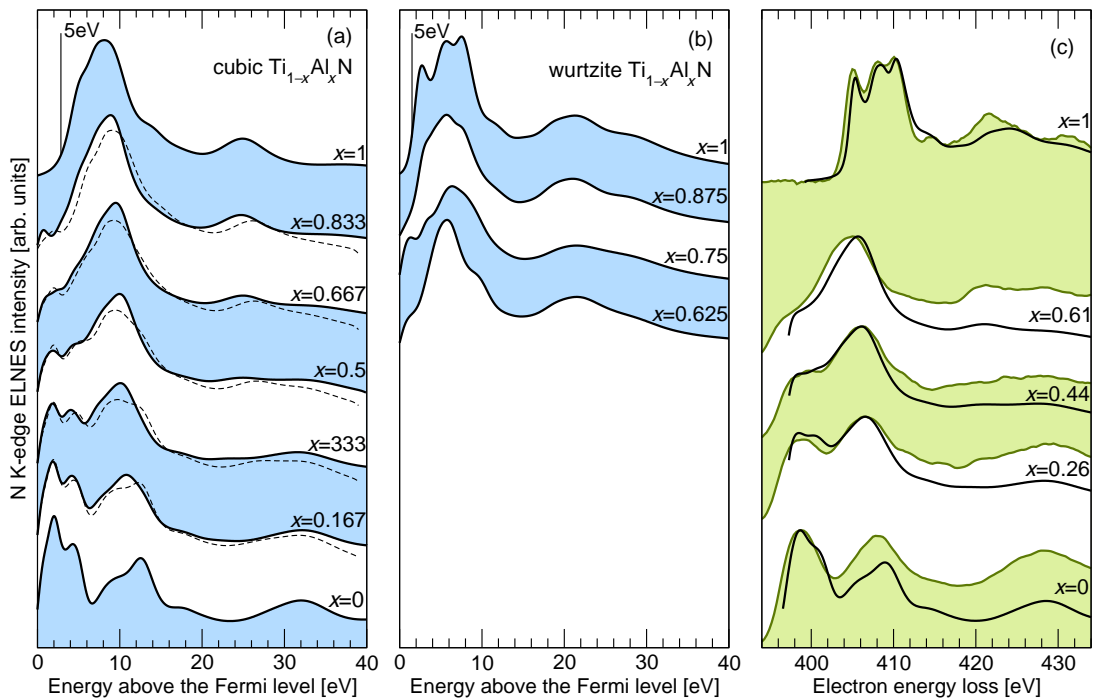}
  \caption{Calculated N K-edge evolution in the (a) cubic and (b) wurtzite phase of Ti$_{1-x}$Al$_x$N alloy. The spectra of pure AlN ($x=1$) are shifted in order to correct for the opened band gap. The dashed line corresponds to a linear interpolation between the binary compounds spectra (see section \ref{sec:shape}). (c) Comparison of the calculated (solid black line) and experimental N K-edge ELNES of Ti$_{1-x}$Al$_x$N. {\color{purple}Note that the structure changes from cubic ($x<0.7$) to wurtzite ($x>0.7$).}}
  \label{fig:TiAlN}
\end{figure*}

The calculated evolution of the N K-edge for Ti$_{1-x}$Al$_x$N is shown by solid lines in Fig.~\ref{fig:TiAlN} for the cubic and wurtzite modifications. The raw ELNES was broadened with a Gaussian having $1\uu{eV}$ FWHM. Moreover, the curves of pure AlN ($x=1$) were shifted along the energy axis (as labelled in Fig.~\ref{fig:TiAlN}) to account for the abrupt change in the Fermi energy %\footnote{As the Fermi energy is taken the energy of the highest occupied state, i.e. the top of the valence band in the case of semiconductors.} 
due to the development of the band gap (no Ti $d$-states present).

The compositional step used for the cubic alloy is $\Delta x=0.167$ (Fig.~\ref{fig:TiAlN}a). Three developments of the main peaks are predicted: (i) the double-maxima at $0$--$5\uu{eV}$ above $E_F$ disappears with increasing AlN content, (ii) the peak at $\approx13\uu{eV}$ for TiN gradually moves to $\approx9\uu{eV}$ for AlN and at the same time its intensity increases, and (iii) a small hump at $\approx32\uu{eV}$ for TiN broadens to an almost undetectable background at $x=0.5$. At the same time, a small hump develops at $\approx26\uu{eV}$ with increasing AlN content. The origins of these composition-induced peak variations can be tracked down to the changes in bonding in the alloy as is discussed later in section \ref{sec:origin_of_peaks}.

The situation is more complicated for the w-Ti$_{1-x}$Al$_x$N alloy (Fig.~\ref{fig:TiAlN}b, compositional step $\Delta x=0.125$). The calculated spectra suggest that the characteristic triple-peak character of the w-AlN N K-edge is levelled out with the addition of only $0.125$ mole fraction of TiN. Additionally, the spectra do not show any clear trends in the peak development as in the cubic case, apart from the peak at $\approx20\uu{eV}$, whose position is not influenced by the composition. This is most likely connected with local relaxations taking place similar to those reported for Nb$_{1-x}$Al$_x$N \citep{holec10}. In particular, N sites near Al tend to have the four-coordinated (wurtzite-like) neighbourhood, while in the vicinity of Ti atoms, five-coordinated local neighbourhoods are preferred (hexagonal \Bk-like). The structure is therefore much more sensitive to the actual arrangement of atoms in the supercell which is reflected, e.g. in much bigger scatter of the ``optimised'' lattice constants for different arrangements of atoms in the SQS (with a constant composition $x$) \citep{holec_inprep}.

To compare the calculated and measured N K-edge evolutions (Fig.~\ref{fig:TiAlN}c), a larger spectrometer broadening parameter of $1.5\uu{eV}$ was used. As a consequence, the double-maximum of the first peak in c-Ti$_{1-x}$Al$_{x}$N at $\approx0$--$5\uu{eV}$ above the $E_F$ ``smears out'' and the measured shape is obtained. It is therefore concluded that the fine double-maximum character is not resolved due to experimental limitations. The experimental spectra were smoothed and normalised to fit the intensity of the highest peak of the simulated pattern in each individual case; no other operation was performed on them. The theoretical spectra were, on the other hand, shifted by the calculated energy of the core-holed core level. The spectra thus obtained exhibit a very good correlation between experiment and theory.

\subsection{Al$_x$Ga$_{1-x}$N alloy}

\begin{figure*}
  \centering
  \includegraphics{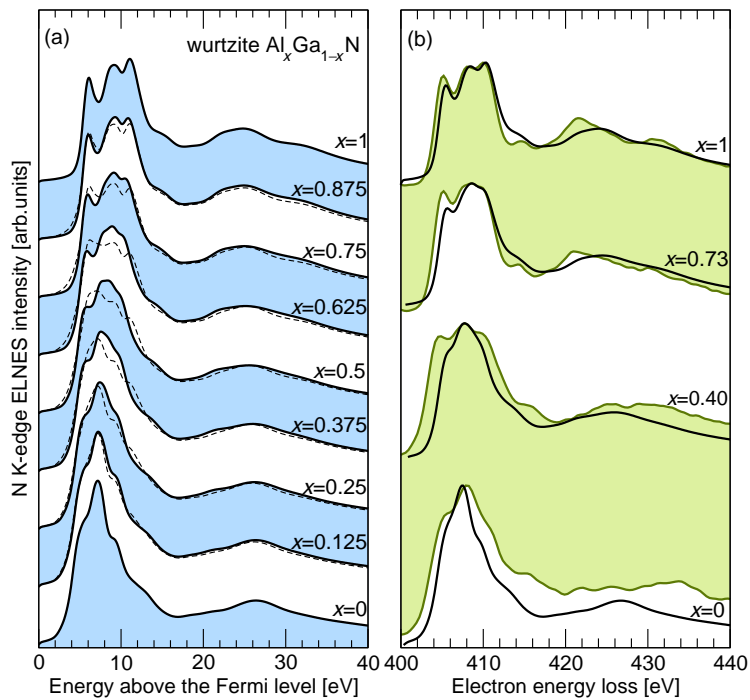}
  \caption{(a) Calculated evolution of N K-edge of the wurtzite Al$_x$Ga$_{1-x}$N alloy and (b) a comparison of calculated (solid black line) with experimental N K-edge ELNES shapes. The dashed lines correspond to linearly interpolated boundary spectra of AlN ($x=1$) and GaN ($x=0$).}
  \label{fig:AlGaN}
\end{figure*}

As another example, the semiconducting wurtzite solid solution of AlN and GaN is chosen to demonstrate the ability of the current approach to predict ELNES. In contrast to the meta-stable Ti$_{1-x}$Al$_x$N, the wurtzite Al$_x$Ga$_{1-x}$N mixture is stable in this modification for all concentrations $x$. Therefore, no local distortions as in the case of the w-Ti$_{1-x}$Al$_x$N alloy are expected, which results in a gradual N K-edge evolution as shown in Fig.~\ref{fig:AlGaN}a. The intensities of the first and third peaks of the  triple-peak shape characteristic for w-AlN decrease with decreasing AlN mole fraction, and become shoulders around a central peak, a shape typical for GaN. The peak at $\approx24\uu{eV}$ above the $E_F$ in the w-AlN spectrum gradually moves to $\approx26\uu{eV}$ for w-GaN with increasing GaN content. The edge onset moves towards the Fermi level reflecting the narrowing band gap from $4.2\uu{eV}$ (AlN) to $1.7\uu{eV}$ (GaN)\footnote{DFT is known to underestimate the values of band gap with respect to the experimental measurements.}. This gradual tranformation of the spectra shape is traced down to the changes in electronic structure, see section \ref{sec:origin_of_peaks}. The predicted evolution of the N K-edge is again confirmed by the experimental observations (Fig.~\ref{fig:AlGaN}b). 

\section{Discussion}

\subsection{Shape and evolution of the N K-edge ELNES}\label{sec:shape}

The spectra of the binary TiN, AlN and GaN systems have been extensively discussed in the literature both from the experimental and theoretical perspective. Despite that, several issues remain unclear, in particular the edge onset energy: {\color{purple}its value varies in the literature within the range of several eVs \citep{serin98,sennour03,mizoguchi04,mackenzie05,holec08,mizoguchi09,mizoguchi09a}. Consequently, we used our own measurements to calibrate the calculations instead of taking spectra from the literature. The lineshapes of binary w-AlN, c-TiN, and w-GaN resemble those previously published for the same materials. 

The edge evolution for c-Ti$_{1-x}$Al$_{x}$N exhibits the same trends as the one previously published by \citet{mackenzie05} (which, however, provided only one intermediate composition). \citet{gago09} used XANES to measure the N K-edge of Ti$_{1-x}$Al$_x$N experimentally. Their XANES spectra have all the main features of our experimental as well as calculated N K-edge ELNES. Also in the case of w-Al$_x$Ga$_{1-x}$N, the calculated evolution of N K-edge ELNES correlates with the here presented experimental data as well with those published previously \citep{radtke04,holec08}\footnote{It is worth noting that the experimental spectra shown by \citet{holec08} do not have subtracted background which, unfortunately, led to incorrect edge shapes and conclusions about the core hole charge.}.}

\citet{mackenzie05} tried to model the edge with a linear interpolation of the boundary binary spectra. Having in mind the problems with the accurate edge onset determination and the lack of cubic AlN (in the B1 structure) for getting a reliable binary spectrum, this approach is questionable. To demonstrate this further, Figs.~\ref{fig:TiAlN}a and \ref{fig:AlGaN}a include the linear interpolations of the binary spectra (dashed lines). Although this may serve as a first (and quick) guess on what the evolution should look like, in many cases the relative intensities and/or positions of the peaks are not predicted correctly.

\citet{craven95} showed using several binary transition metal nitrides (TMN) that the spacing between the double-maximum peaks increases with increasing lattice parameter. This is not predicted for the c-Ti$_{1-x}$Al$_x$N alloy (Fig.~\ref{fig:TiAlN}) where the lattice parameter decreases from $4.25\uu{\mbox{\AA}}$ for TiN ($x=0$) to $4.07\uu{\mbox{\AA}}$ for AlN ($x=1$) \citep{mayrhofer06a}, but the peak spacing is almost unaffected. The reason for this is that the bonding of various TMN is similar and the peaks follow small shifts of the density of states associated with the varying valence configuration. On the contrary, the peak shifts in the Ti$_{1-x}$Al$_x$N evolution result in the first place from the changing character of bonding (see section \ref{sec:origin_of_peaks}). Consequently, when the ELNES of the two boundary binary systems are similar, the simple approach of interpolating between the binary ELNES spectra \citep{mackenzie05} is expected to give acceptable results, see e.g. Ti$_{1-x}$V$_x$N \citep{mackenzie97} or In$_{1-x}$Ga$_x$N \citep{keast02}. The extremely small cells (1 In and 1 Ga atom for In$_{0.5}$Ga$_{0.5}$N) used in the latter reference satisfactorily reproduced the N K-edge evolution, and a much more computationally expensive approach (using the averaging of several N core-holed sites in SQSs as in the present paper) employed by \citet{holec10-diss} is not necessary.

On the other hand, when the spectrum evolution is pronounced, small ordered structures do not provide reliable predictions. This has been shown by \citet{mackenzie05} for the case of c-Ti$_{1-x}$Al$_x$N and by \citet{holec08} for w-Al$_x$Ga$_{1-x}$N. In such cases, the approach adopted here is essential.

\subsection{Electronic origin of the peaks}\label{sec:origin_of_peaks}

There exists extensive literature on the origin of peaks for semiconducting III-N binaries. As summarised by \citet{mizoguchi09a} using the overlap population analysis, the main peak structure (up to $\approx10\uu{eV}$ above the edge onset) reflects the anti-bonding N--cation interaction while the later peak ($20$--$30\uu{eV}$ above the edge onset) corresponds to cation--cation (mostly anti-bonding) interactions. The difference between the AlN and GaN ELNES shapes can be traced down to the presence of the valence $d$-electrons in GaN which cause (slight) redistribution of the valence density of states, and consequently also the unoccupied density of states. The electronic structure of the valence band of InN is similar to that of GaN thus resulting in a similar ELNES spectrum.

\begin{figure*}
  \centering
  \includegraphics{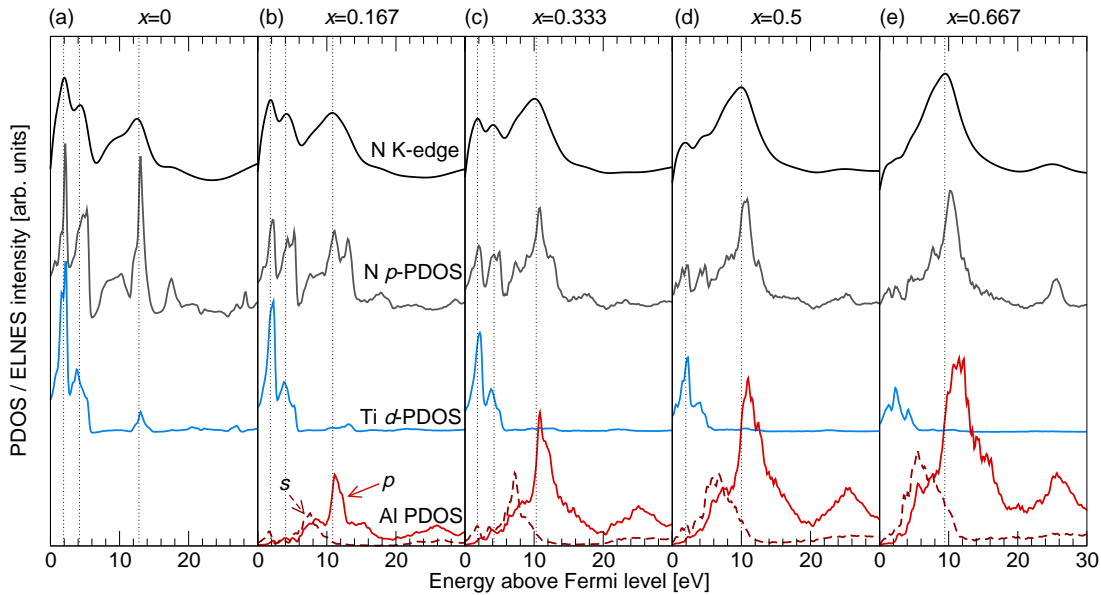}
  \caption{PDOS/ELNES spectra for c-Ti$_{1-x}$Al$_x$N with (a) $x=0$, (b) $x=0.167$, (c) $x=0.333$, (d) $x=0.5$, and (e) $x=0.667$. The peaks in N K-edge ELNES closely reflect the distribution of the unoccupied N $p$ states. The overlap of the individual PDOSes suggest that the double peak at $\approx2$--$4\uu{eV}$ above $E_F$ arises from hybridisation of N $p$ and Ti $d$ states while the peak at $\approx10\uu{eV}$ can be ascribed primarily to N $p$ states hybridised with Al $p$ states.}
  \label{fig:TiAlN_el}
\end{figure*}

The ground state projected density of states (PDOS) in Fig.~\ref{fig:TiAlN_el} helps to understand the meaning and evolution of peaks in ELNES of the c-Ti$_{1-x}$Al$_x$N alloy. The site and symmetry projected DOS were obtained by averaging the corresponding quantities over all sites occupied with the same specie. The final states of the N K-edge transition are unoccupied N $p$-states which clearly correlate  with the ELNES. Based on the PDOS overlaps it can be concluded that the double-maximum structure just above the edge onset arises from the N $p$--Ti $d$-states interaction which agrees with the findings of \citet{tsujimoto05} and \citet{lazar08} for binary c-TiN. The second peak at around $10$--$12\uu{eV}$ above $E_F$ have the strongest contribution from the N $p$--Al $p$-states interaction, the only exception being pure TiN (no Al present) where a small peak in Ti $d$-states at the same position can be detected. The different interactions contributing to this delayed peak are responsible for a sharper maximum with clear shoulders in the case of pure TiN while resulting in a rather broad (and nearly symmetric) shape when Al is present, see Fig.~\ref{fig:TiAlN_el}. The peak position changes by almost $1\uu{eV}$ upon adding $x=0.167$ mole fraction of AlN to TiN, while further increase of AlN content results in only a small shift of the peak ($0.5\uu{eV}$ for increasing $x$ from $0.167$ to $0.667$). This can serve as an example why the simple interpolation between the properties of binary compounds (as suggested by \citet{mackenzie05}) does not work.

\begin{figure}
  \centering
  \includegraphics{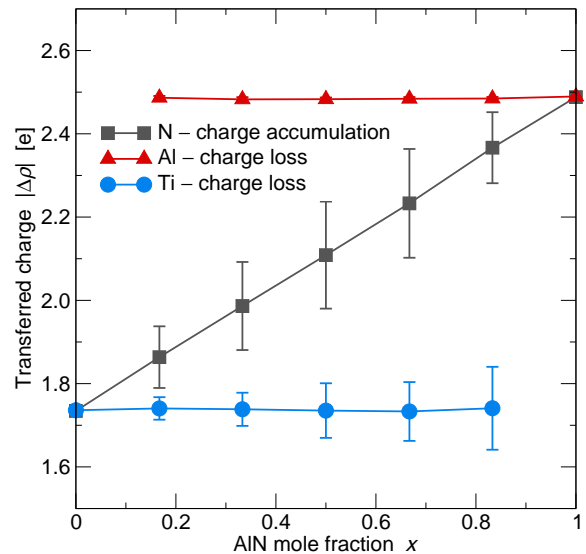}
  \caption{Bader analysis of a charge transfer (absolute values) on atoms of individual species.}
  \label{fig:bader}
\end{figure}

The bonding of the c-Ti$_{1-x}$Al$_x$N alloy consists of a mixture of covalent and ionic type. The bonding of TiN has been discussed many times in the literature \citep[see, e.g. a review by][]{schwarz87}. It was concluded that the covalent part is realised by the $sp^3d^2$ orbitals (interaction of Ti $4s$ and $3d$-states with N $2p$-states). Additionally, the interaction between Ti $3d$ orbitals with the $t_{2g}$ symmetry causes a non-zero DOS at the Fermi level resulting in the metallic character of the compound. The influence of Al on bonding in the alloy was discussed by several authors \citep{mayrhofer06a,alling07,rovere10}, generally showing a gradual weakening of the $sp^3d^2$ hybridisation (which is reflected by the decreased intensity of the first double-maximum peak in the N K-edge ELNES, see Figs.~\ref{fig:TiAlN} and \ref{fig:TiAlN_el}). Additionally, the Bader analysis \citep{bader90} as implemented in Wien2k shows that there is a significantly increased charge transfer from metallic sites to N resulting in a stronger ionic bonding with increasing AlN mole fraction (see Fig.~\ref{fig:bader}). It is interesting to note that the different Al sites ``provide'' on average always almost the same charge to be transferred on N sites (practically no scatter around the mean values shown by the triangles in Fig.~\ref{fig:bader}), but the charge transferred from Ti sites is much more influenced by the alloy composition. This is likely to be due to different degrees of hybridisation between Ti and N atoms depending on the actual neighbourhood of N atoms (i.e. second-order neighbours of Ti sites).

\subsection{Influence of the local environment}

\begin{figure*}
  \centering
  \includegraphics{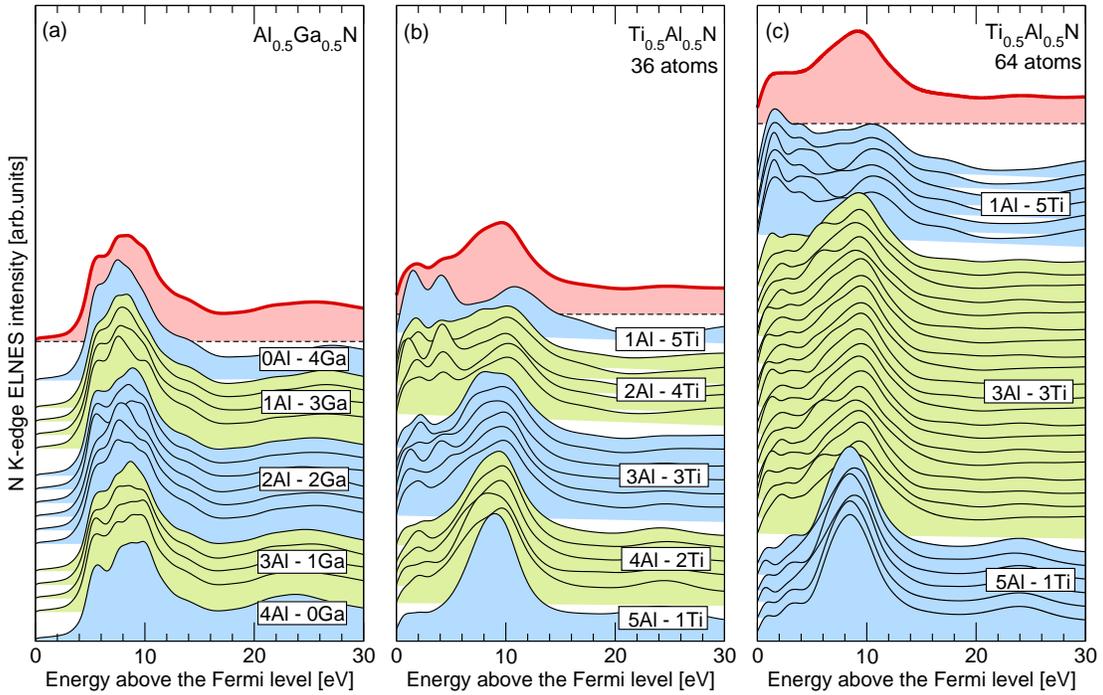}
  \caption{{\color{purple}The resulting spectra representing the alloys (thick curves on top of each panel) and spectra calculated by placing the core hole on individual N sites and sorted accorting to the nearest neighbour configurations.} (a) Wurtzite Al$_{0.5}$Ga$_{0.5}$N, (b) cubic Ti$_{0.5}$Al$_{0.5}$N supercells with 36 atoms, and (c) cubic Ti$_{0.5}$Al$_{0.5}$N supercells with 64 atoms.}
  \label{fig:loc_env}
\end{figure*}

There is some controversy in the literature on how big the supercells should be in order to suppress the mutual interactions between core holes. For example, \citet{mizoguchi04} and \citet{tanaka09} claimed that cells with more than $100$ atoms are needed while $32$ atom cells were found sufficient by \citet{lazar04} for GaN and by \citet{holec08} for w-AlN. To address this issue we plotted the individual spectra resulting from core hole being placed on various N sites, and sorted them according to the number of the nearest neighbours of each specie (in total 4 for the tetrahedrally coordinated wurtzite structure (Fig.~\ref{fig:loc_env}a) and 6 for the octahedrally coordinated cubic structure (Fig.~\ref{fig:loc_env}b,c)) surrounding the particular N site with the core hole. {\color{purple}The numbers of spectra corresponding to individual local environments (i.e. the nearest neighbour configuration) results from their real counts in the used supercells. Although configurations of the nearest neighbours of some N sites are the same, the higher-order nearest neighbour configurations differ which is why small variations between individual spectra labelled with the same local environment are obtained. The thick lines on top of each panels in Fig.~\ref{fig:loc_env} were obtained by averaging all the curves underneath (their number is the same as the number of N sites in the supercell), and thus account for the statistical distribution of various local environments of N atoms.}

The graphs clearly demonstrate the huge differences between spectra depending on the local environment of the N site where the excitation takes place. At the same time one can see that almost doubling the number of atoms in the supercell (from 36 in Fig.~\ref{fig:loc_env}b to 64 in Fig.~\ref{fig:loc_env}c) does not alter the N K-edge significantly. The small changes might be due to  having insufficiently big cell in the case of 36 atoms, but also could be due to a non-representative (i.e. not SQS-like) cell for the bigger structure which is suggested, e.g. by not having the 2Al, 4Ti local environment present. In summary, Fig.~\ref{fig:loc_env} demonstrates that (i) the local environment influences the final shape of the edge much more drastically than the actual cell size (provided the cell is big enough to model the ``randomness'' of an alloy), and (ii) the similarity of the curves from Al-, Ti-, Ga-rich local neighbourhoods to those of pure AlN, TiN and GaN, respectively, gives some grounds for the interpolation approach (see section \ref{sec:shape}).

\subsection{Energy of the edge onset}

The edge onset energy is another important feature of the edge; for example, \citet{mizoguchi03} predicted a range of $\approx2\uu{eV}$ for the N K-edge onset of AlN depending on the crystal structure thus having a potential to distinguish between these allotropes. It is, however, not straightforward to define the edge energy, in particular due to the {\color{purple}ambiguity} in the background subtraction as well as due to the background noise itself. Consequently, we have chosen the energy of the first inflection point above the edge threshold for the comparison between experiment and theory (the edge onset is about $2$--$4\uu{eV}$ below).% This is shown in Fig.~\ref{fig:AlGaN_edge-onset} for the Al$_x$Ga$_{1-x}$N alloy.

% \begin{figure}
%   \centering
%   \includegraphics{figs/edge_onset}
%   \caption{The energies of the calculated (full circles) and measured (asterisks) N K-edges of Al$_x$Ga$_{1-x}$N. The values correspond to the position of the first inflexion point after the edge onset (see text for explanation).}
%   \label{fig:AlGaN_edge-onset}
% \end{figure}

It is not surprising that we obtained excellent agreement for the binary systems ($<0.1\uu{eV}$ for GaN and $\approx0.35\uu{eV}$ for AlN) since for these systems the edge onset energy was used as a fitting parameter for the core hole charge (see section \ref{binaries}). However, both the experiment and the theory suggest that the energy of the inflection point does not vary too much with the composition. The variations within $0.2\uu{eV}$ (theory) and $0.4\uu{eV}$ (experiment) can be regarded as the accuracy of the present approach (due to, e.g. the background subtraction on the experimental side or the supercell design/size on the theoretical side).

\begin{figure*}
  \centering
  \includegraphics[scale=0.7]{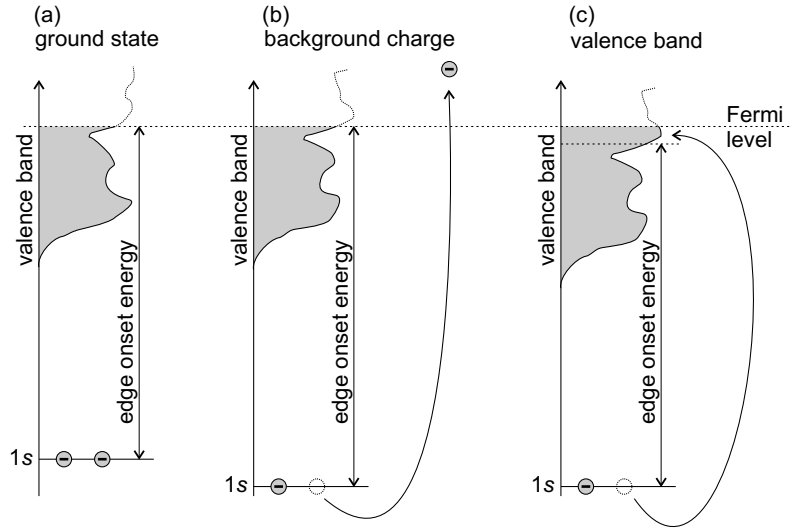}
  \caption{(a) A ground state and an excited state with a full core hole: (b) the excited electron is put as the ``background charge'' (i.e. effectively removed from the system) while (c) it is put in the valence band (into prior the excitation unoccupied states). The scheme corresponds to a metallic system.}
  \label{fig:excitation}
\end{figure*}

A common method calculating the excitation energy is to calculate the difference between total energies of the initial ground state and the final state with a full core hole \citep[see for example][]{mizoguchi03, rashkova07, tanaka09} which, in principle, follows the excitation process\footnote{Still, this is not fully justified as DFT is a {\em ground state} theory (Fig.~\ref{fig:excitation}a).}. This results in values of $384.5\uu{eV}$ and $368.2\uu{eV}$ for w-AlN and c-TiN, respectively, which are hugely underestimated as compared with the experimental values of $402\uu{eV}$ and $397\uu{eV}$. The reason for this is that the excited electron was put as the background charge (Fig.~\ref{fig:excitation}b) rather than in the unoccupied states (Fig.~\ref{fig:excitation}c). When excited to the unoccupied states values of $406.4\uu{eV}$ and $404.6\uu{eV}$ for the w-AlN and c-TiN are obtained which are much closer to the experimental values. \citet{rashkova07} showed that a further improvement (towards the experimental values) could be obtained by performing spin polarised calculations. Nevertheless, the edge shapes as well as the energies of the initial core levels are almost identical using both approaches (background charge vs. valence band) thus yielding comparable results (except for the edge onset), which is in agreement with \citet{hebert07}.

When the edge onset is calculated as the total energy difference between the ground and full core hole states, its value is a given number without any degree of freedom for adjustments. The corresponding ELNES shape then should be calculated with exactly $0.5\uu{e}$ core hole \citep{paxton00}. This could be useful when estimating, e.g. ELNES of experimentally inaccessible phases. However, when one uses the core hole charge as a fitting parameter (as in this paper) then it is well justified that also the edge onset is not a unique number but instead a function of the core hole charge. For evaluation of the energy difference between the core state  and the lowest unoccupied state, however, the approach with background charge is more appropriate since, for example, in the case of the conductive TiN it allows to get the energy of the lowest (originally) unoccupied state (compare Figs.~\ref{fig:excitation}b and c). %This is not a problem for semiconducting AlN where the lowest (originally) unoccupied state can be estimated even for the core hole charge put in the conduction band as the excited (part of the) electron is the only charge occupying this band.

\section{Conclusions}

In this paper we have demonstrated a semi-empirical approach for predictive calculations of the N K-edge ELNES of various classes of alloys (cubic vs. wurtzite, metallic vs. semiconducting). We used fractional core holes with charges, carefully adjusted according to the edge shape and the onset energy, to reproduce experimental spectra. Subsequently we utilised these to model the ELNES spectra of alloys using the special quasi-random supercells. We introduced core holes on all individual N sites, and by averaging these spectra we achieved a representative alloy spectrum. A comparison with the experimental measurements (for c-Ti$_{1-x}$Al$_x$N and w-Al$_x$Ga$_{1-x}$N systems) yielded an excellent agreement on both the edge shapes (including peak positions and relative intensities) as well as the edge onset energies. Finally, we related the individual peaks in the N K-edge ELNES to various interactions between cations and N atoms demonstrating that the decrease in intensity of the N K-edge structure $\approx3\uu{eV}$ above the edge onset reflects a weakening of the $sp^3d^2$ hybridisation with increasing Al content in Ti$_{1-x}$Al$_x$N.

\section{Acknowledgements}

Financial support by the START Program (Y371) of the Austrian Science Fund (FWF) and by the UK Engineering and Physical Sciences Research Council (EPSRC) is greatly acknowledged.

% \bibliographystyle{elsarticle-harv}
% \bibliography{/home/david/work/papers/codes.bib,ELNES_ternary.bib,/home/david/work/papers/nitride_coatings.bib,/home/david/work/papers/focus_papers/EELS/EELS.bib}
% \bibliography{codes.bib,ELNES_ternary.bib,nitride_coatings.bib,EELS.bib}


\end{document}